\documentclass[preprint,twocolumn,10pt,a4paper,byrevtex,prl,superscriptaddress,showpacs]{revtex4-1}
\usepackage{natbib}
\usepackage{graphicx}
\usepackage{amsfonts}
\usepackage{amsmath}
\usepackage{amssymb}
\usepackage{color}

\begin{document} 

\title{Magnetic state dependent transient lateral photovoltaic effect in patterned ferromagnetic metal-oxide-semiconductor films}

\author{Isidoro Martinez}
\affiliation{Dpto. Fisica de la Materia Condensada C-III, Instituto Nicolas Cabrera (INC) and Condensed Matter
Physics Institute (IFIMAC), Universidad Autonoma de Madrid, Madrid 28049, Spain}

\author{Juan Pedro Cascales}
\affiliation{Dpto. Fisica de la Materia Condensada C-III, Instituto Nicolas Cabrera (INC) and Condensed Matter
Physics Institute (IFIMAC), Universidad Autonoma de Madrid, Madrid 28049, Spain}

\author{Antonio Lara}
\affiliation{Dpto. Fisica de la Materia Condensada C-III, Instituto Nicolas Cabrera (INC) and Condensed Matter
Physics Institute (IFIMAC), Universidad Autonoma de Madrid, Madrid 28049, Spain}

\author{Pablo Andres}
\affiliation{Dpto. Fisica de la Materia Condensada C-III, Instituto Nicolas Cabrera (INC) and Condensed Matter
Physics Institute (IFIMAC), Universidad Autonoma de Madrid, Madrid 28049, Spain}

\author{Farkhad G. Aliev}
\affiliation{Dpto. Fisica de la Materia Condensada C-III, Instituto Nicolas Cabrera (INC) and Condensed Matter
Physics Institute (IFIMAC), Universidad Autonoma de Madrid, Madrid 28049, Spain}

\begin{abstract}

We investigate the influence of an external magnetic field on the magnitude and dephasing 
of the transient lateral photovoltaic effect (T-LPE) in lithographically 
patterned Co lines of widths of a few microns grown over naturally passivated p-type Si(100).
The T-LPE peak-to-peak magnitude and dephasing, measured by lock-in or through the characteristic
time of laser OFF exponential relaxation, exhibit a notable influence of the magnetization direction of the ferromagnetic overlayer.
We show experimentally and by numerical simulations that the T-LPE magnitude 
is determined by the Co anisotropic magnetoresistance. On the other hand, 
the magnetic field dependence of the dephasing could be described by the influence 
of the Lorentz force acting perpendiculary to both 
the Co magnetization and the photocarrier drift directions. 
Our findings could stimulate the development of fast position sensitive detectors 
with magnetically tuned magnitude and phase responses.

\end{abstract}

\maketitle


The non-uniform illumination of a semiconductor surface or ultrathin metallic films forming 
Metal-Oxide-Semiconductor (MOS) structures generates an electric 
field parallel to the surface or to the Schottky barrier due to a photocarrier drift in 
non-equilibrium conditions. The straight-forward way to maximize the potential 
difference is to generate photo-carriers effectively (by using a laser with its photon 
energy exceeding the gap) with a spot diameter much less than the distance between the asymmetrically
situated lateral contacts. This potential difference is known as the lateral photovoltaic effect (LPE) 
\cite{WALLMARK1957,Lucovsky1960,Boeringer1994,Zhao2006}. On the practical side, the LPE has been widely 
used to develop high precision position-sensitive  detectors (PSD) \cite{Niu1987,Yu2010,Jin2013}. 
The recent interest in Co/SiO$_{2}$/Si structures has been related to their high PSD sensitivity for the visible, 
ultraviolet or infrared range \cite{Xiao2007,Kong2008,Qi2010} by adjusting the Co thickness. 

Advances in electron beam lithography have permitted the development of MOS structures where the LPE 
can be investigated along patterned, few micron wide, metallic line structures.  Recently Cascales et al. 
\cite{Cascales2014} have found that the time dependent photovoltaic response along such structures is 
different from the one observed in the wide LPE devices. Specifically, peak-like transitorials which 
present a sign inversion of the T-LPE 
in the laser OFF state followed by a nearly exponential relaxation back to equilibrium have been qualitatively 
explained with a simple model taking into account not only resistance and capacitance, but also the 
local inductance of the metallic line structure deposited on top of a Schottky barrier.

Here we investigate the dependence on the magnetic state of both the magnitude and the phase 
response of the T-LPE, studied by a lock-in technique and by analyzing the relaxation back to equilibrium 
of the T-LPE time trace, in lithographically patterned (21 nm thick, 10$\mu m$ wide and 1500 $\mu m$ long) Co lines. 
The structures were deposited on a naturally passivated (about 2 nm SiO$_{2}$) Silicon (100) substrate. 
More details on the preparation and characterization of samples may be found in \cite{Brems2007,Herranz2009,Cascales2014}. 

We have observed and qualitatively explained the substantial influence of the magnetic state of the Cobalt 
ovelayer on the magnitude and phase of T-LPE. On one hand, numerical simulations and experiments confirm a direct 
link between the field dependent T-LPE and the anisotropic magnetoresistance. On the other hand, we observe a signal
dephasing dependent on the magnetization direction, which could be linked to the inversion of the direction of the Lorentz
force acting on the photocarriers drifting along the metallic line structure.


We have investigated the magnetic field dependent T-LPE with two electronically 
different setups. The first system obtains a direct
measurement of the T-LPE characteristics with microsecond resolution and was described elsewhere\cite{Cascales2014}. 
A second measurement scheme employing the 
lock-in technique was used to observe the influence of an external magnetic field 
on the T-LPE amplitude as well as the phase difference between the reference 
signal and the T-LPE response at the first harmonic. Such lock-in measurements, with 
laser pulses modulated by a reference signal in the kHz range, have been previously 
used to characterize the lateral photovoltaic effect in MOS-type PSDs \cite{Niu1987}. 
The phase of the photovoltage has been shown as alternative, high precision 
detection method of the position of a signal\cite{Niu1989}. The main advantage
of such a method is related with its insensitivity to the variation of the incident light intensity. 
Here we use a similar techique which additionally incorporates the possibility 
to determine LPE phase changes as a function the external magnetic field.

\begin{figure}[!h]
 \begin{center}
  \includegraphics[width=\columnwidth]{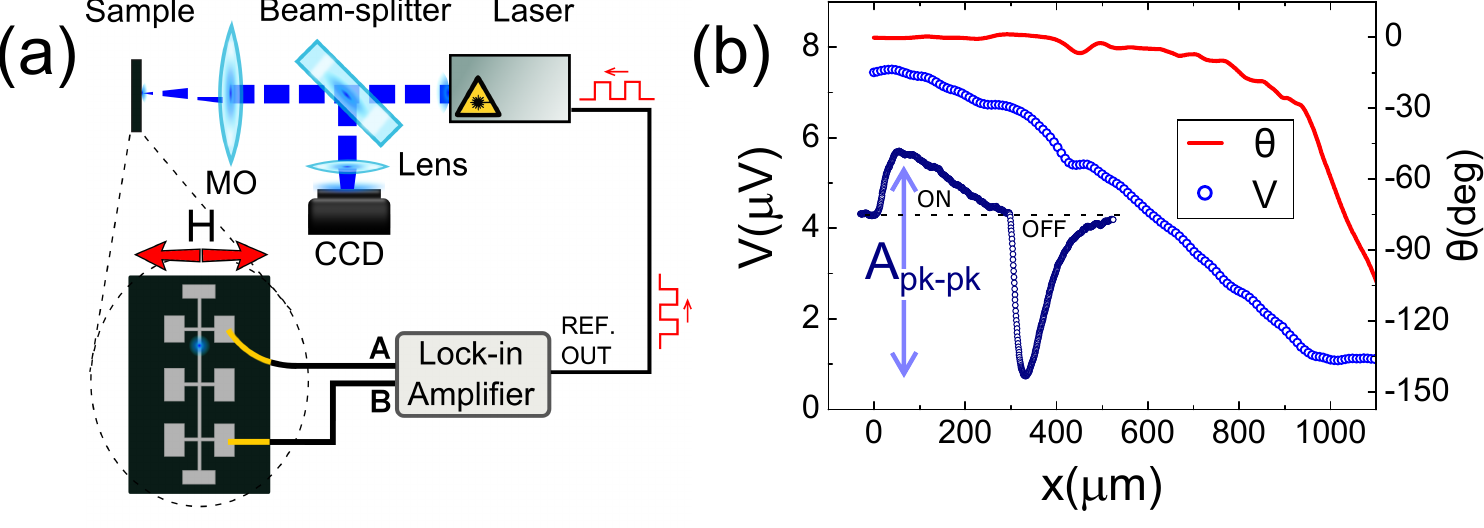}
  \caption{(a) Diagram of the lock-in experimental setup. (b) Voltage and phase of the LPE signal registered by the lock-in
for different laser spot positions for a laser power of 2.8 mW. The inset presents a 
typical T-LPE response when the laser is switched ON/OFF.}\label{fig:fig1}
 \end{center}
\end{figure}

The optical setup used for the measurements is schematically shown in Fig. \ref{fig:fig1}(a). The laser beam is focused
onto the sample by a microscope objective lens (MO) (50x, 0.42 NA, Plan APO, working distance 21 mm) and
the image of the sample is relayed into a CCD
camera by using the objective and a beam splitter. The spot size has a diameter of around 2 $\mu m$.
The potential difference
created along the line is measured by contacting three pairs of $500\times 500$ $\mu m^{2}$ 
Cobalt pads with gold wires and using indium. The T-LPE has been studied by applying 
a train of periodic laser beam pulses, with a power between 0.7-3 mW
which are modulated by a reference signal (typically of 787 Hz). The electronic setup which obtains
the time trace of the transient LPE response was described elsewhere\cite{Cascales2014}. The second experimental
technique consists on feeding the LPE voltage as the input into a lock-in so the magnitude and phase
of the voltage at the reference frequency are registered. 
A TOPTICA-iBeam Smart diode laser which emits light of 487 nm
of wavelength $\lambda$ has been used. The optical setup is common to both electronic
detection methods. An in-plane magnetic field perpendicular to the Cobalt line
is applied by two Helmholtz coils as is schematically shown in Fig. \ref{fig:fig1}(a).

A typical T-LPE response when the device is subjected
to a periodically modulated light beam shows peak-like transitorials presenting a 
sign inversion in the OFF state (see the inset of Fig. \ref{fig:fig1}(b)). The OFF state peak is followed by a nearly exponential relaxation
back to equilibrium. The peak-to-peak amplitude ($A_{pk-pk}$ and the characteristic time $\tau_{OFF}$ of the laser OFF exponential relaxation will
be discussed further below.

Let us now discuss the analysis of the lock-in measurements. As can be seen in
the main part of Fig. \ref{fig:fig1}(b), the magnitude depends linearly
on the position of the laser spot (typical LPE behaviour). If we take the phase $\theta$
at one end of the device ($x=0$) as the reference, then the phase of the
signal can be seen to deviate from 0 as the laser spot moves away from the positive
electrode. This change in phase is due to the change in distance that the signal 
has to travel to reach the electrode as the laser spot is moved. Even though our sample has a total length of 1.5 mm, 
the dependence of the phase on the position presents a similar
behaviour as in Ref.\cite{Niu1987} (where the sample size is of the order of a cm),
although the total phase change of around 60 $^\circ$ is somewhat smaller.

\begin{figure}[!h]
 \begin{center}
  \includegraphics[width=\columnwidth]{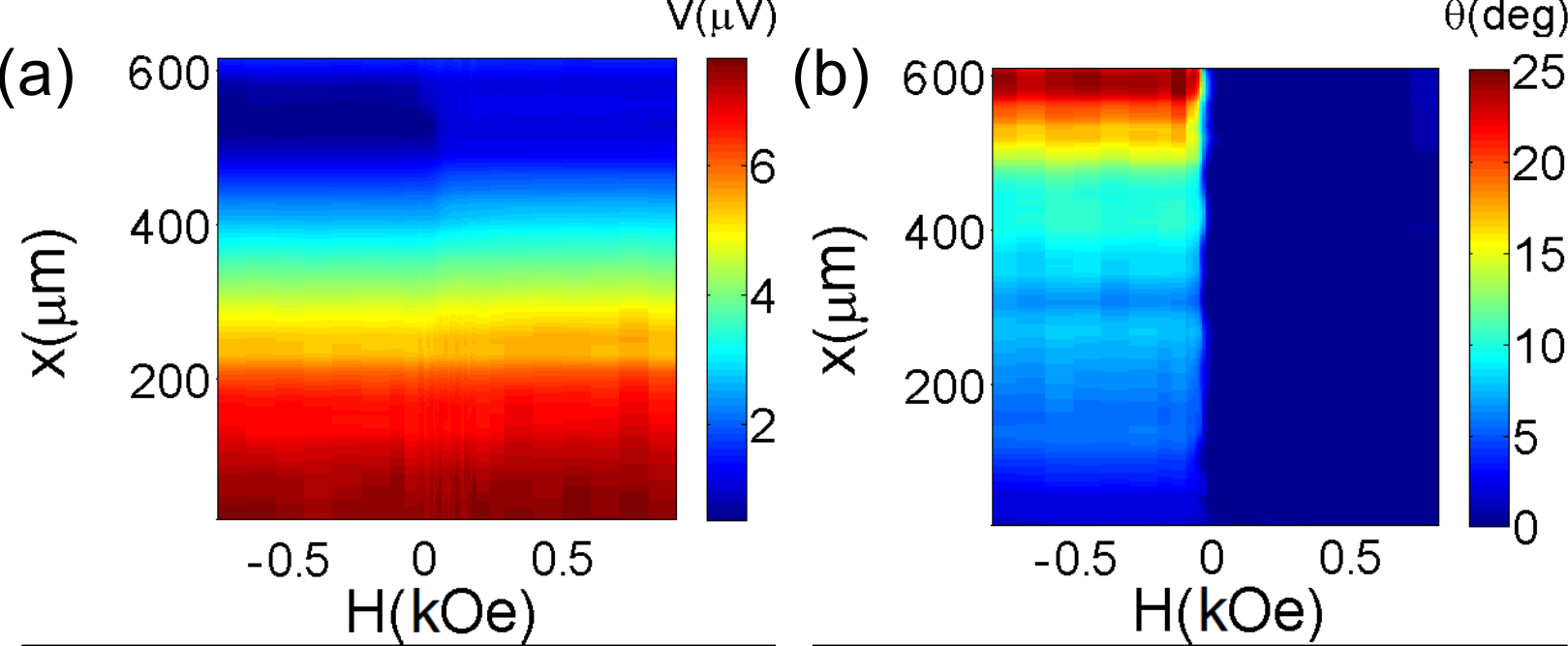}
  \caption{(a) Amplitude and (b) phase of the signal registered by the lock-in as a function of the distance from one
  end of the sample. The phase at the saturation field
  has been subtracted for every spot position for clarity.}\label{fig:fig2}
 \end{center}
\end{figure}

We next discuss the dependence of the voltage and phase on the position in the presence
of an external magnetic field. As can be seen in Fig. \ref{fig:fig2}(a) and (b), the voltage amplitude
presents a close to linear dependence on the position (consistent with the T-LPE) while an abrupt change in the 
phase occurs for fields close to the field where the Co lines exhibit an AMR peak (see below).
We have subtracted the phase values for every position for the positive saturation field (right row of the
plots), in order to observe only the variation with the field. 
This behaviour has been consistently observed in our samples.

As can be seen using the lock-in technique, we find the change in magnitude due to the 
magnetic field being less than 3\%. Therefore, we used a more sensitive 
T-LPE measurement technique to determine  with higher precision
the influence of the magnetic state of the Co lines the possible variation of both magnitude and 
characteristic relaxation back to equilibrium as a 
function of the magnetic state of Co lines.

\begin{figure}[!h]
 \begin{center}
  \includegraphics[width=\columnwidth]{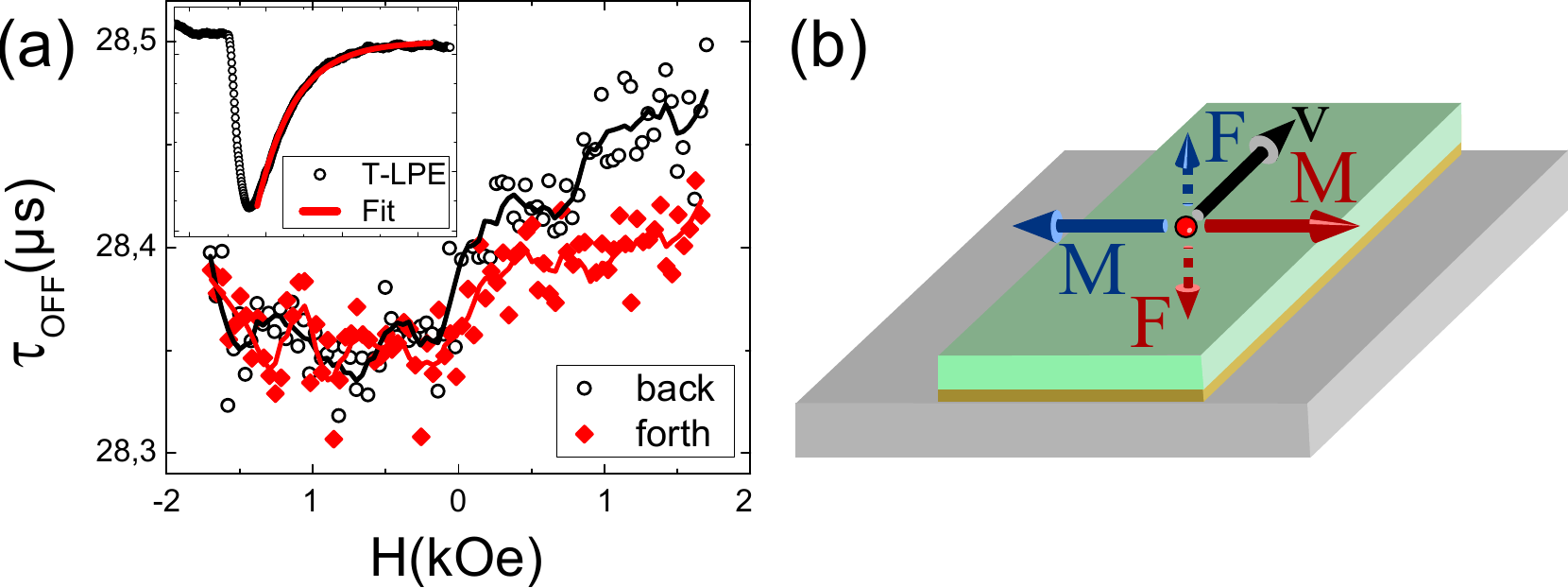}
  \caption{(a) Dependence of the laser OFF relaxation time on the external magnetic field for a 10 $\mu m$ Co line. 
  The inset presents the exponential fitting done to estimate $\tau_{OFF}$.
(b) Diagram representing change in direction of the Lorentz force acting over photocarriers 
due to an inversion of the magnetization of the ferromagnet.}\label{fig:fig3}
 \end{center}
\end{figure}

The variation of an external magnetic field has been also found to influence both the magnitude $A_{pk-pk}$ and
the relaxation time $\tau_{OFF}$ of the laser OFF transition. 
Figure \ref{fig:fig3}(a) presents an analysis of $\tau_{OFF}$ $vs.$ $H$ for a full
field cycle. The relaxation time $\tau_{OFF}$ was obtained by fitting an exponentially decaying
function to the decaying part of the OFF peak (inset of Fig. \ref{fig:fig3}(a)).
As can be seen, the relaxation time for negative fields is noticeably
lower than for positive fields, and a step-like transition between both regimes is
seen close to the switching field of the Co line, known from AMR measurements.
We should note that the branches do not meet at the starting point (positive saturation field).
This is due to the fact that the measurements take a considerable amount of time to carry
out (a few hours), and the temperature conditions could slightly change during the measurement.

The experimentally observed variation of the T-LPE dephasing with the direction of the magnetization 
may be qualitatively explained by the toy model shown 
in Fig. \ref{fig:fig3}(b). For a fixed direction of the velocity of the 
charge carriers, depending on the direction of the applied field, the resulting Lorentz force
may shift the carriers upwards or downwards. If the magnetic field forces the carriers downwards, this will
promote the recombination processes, accelerating the decay of the laser OFF peak. Conversely,
if the recombination process is delayed because the carriers remain in the Co for a longer time,
$\tau_{OFF}$ will increase.

\begin{figure}[!h]
  \begin{center}
  \includegraphics[width=\columnwidth]{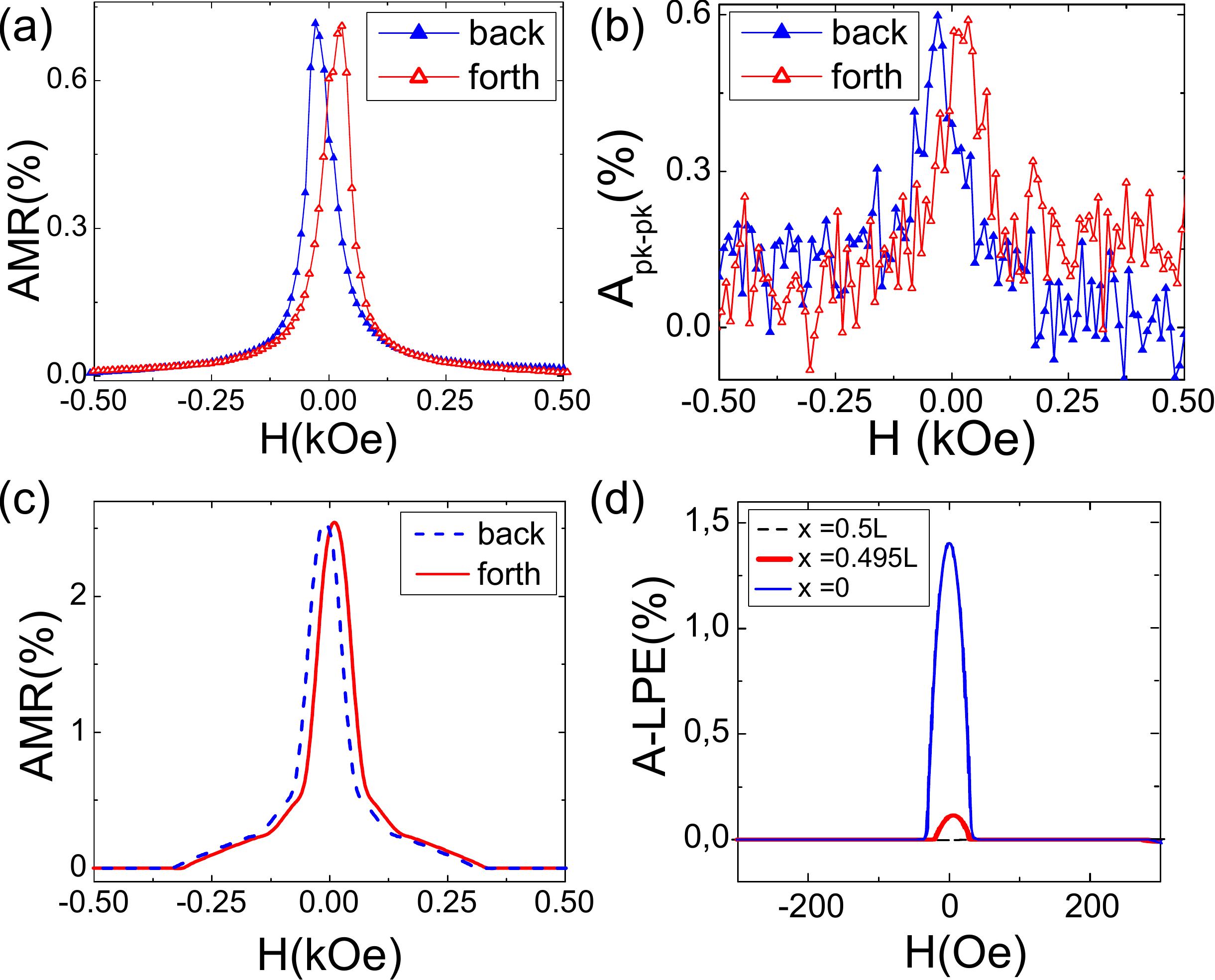}
  \end{center}
  \caption{Comparison of the (a) anisotropic magnetoresistance and
  (b) anisotropic lateral photovoltaic effect for a 10$\mu$m  wide Cobalt line. (c) Simulated AMR of a stripline with field applied perpendicular to the line. 
  (d) Simulated increase in the LPE potential generated by the laser due to the AMR for different positions of the laser spot.}
  \label{fig:fig4}
  \end{figure}

As a further step to clarify the possible influence of the external magnetic field on the T-LPE magnitude,
we carried out magnetoresistance measurements, with the applied field in plane and 
perpendicular to the Co line. Both experiments are done by cycling the field from a positive saturation value down to 
a negative saturation value (back branch), and back to the positive saturation field (forth branch).
Resistance measurements as a function of the field reveal anisotropic magnetoresistance (AMR) effects.
The AMR of the Cobalt line was found to present a peak at low fields (around 20-30 Oe), 
with a maximum AMR ratio of 0.6\% as can be seen Fig \ref{fig:fig4}(a). 
This is a well-known effect in ferromagnets when the local magnetization interacts with 
an electron current, depending on the relative angle between
the local magnetization and the direction of the current (see the Supplemental Material \cite{suppmat}).
  
We have found that the T-LPE
amplitude is enhanced by 0.6\% (see Fig \ref{fig:fig4}(b)), at fields which correspond to the maximum in AMR of
the Co line devices. We define this effect as Anisotropic Transient Lateral Photovoltaic Effect (A-T-LPE). 
If the MOS system is coated with gold, the absolute values of the A-T-LPE are reduced 
almost by half due to the increment of absorbed light and increment of the conductance 
(see the Supplemental Material\cite{suppmat}). 
These observations indicate that the T-LPE peak-to-peak amplitude variation with an applied magnetic field is 
directly related with anisotropic magnetoresistance.

We support this conclusion by micromagnetic simulations and numerical simulations. 
In Figure \ref{fig:fig4}(c) a calculated magnetoresistance curve is shown, indicating the same qualitative 
behavior as the measured one. A larger remanence can be achieved in the simulations by 
artificially introducing disorder (see Supplemental Material\cite{suppmat}). For a second type of simulation, 
the induced voltage created by photocurrents moving away from the laser spot is considered for 
different positions $x$ with respect to one of the contacts. Fig. \ref{fig:fig4}(d) shows this fact for the case of a laser excitation 
at the same distance of each contact (black dashed line, zero signal) and at different distances 
from each contact (non-zero response), expressed in units of the line length $L$. All the non-zero signals are found at low fields, when 
magnetoresistance is high, due to the presence of domains in the unsaturated sample, just like 
in measurements. This suggests the strong influence of magnetoresistance in the signal measured as a function of field.
Further explanations of the simulation methods can be found in the Supplemental Material \cite{suppmat}.


To sum up, we have studied the influence of the magnetic state of ferromagnetic overlayer on the 
transient lateral photovoltaic effect in 10 $\mu$m wide Co/SiO$_{2}$/Si line structures. Two different 
experimental methods, (i) a direct time dependent response with microsecond resolution and (ii) a
lock-in technique, have been used to investigate the possible changes in dephasing of the 
T-LPE with magnetization direction. For the lock-in measurements, we compare the LPE magnitude and phase in each spatial point along the Co 
line structure at different values of the external field with the 
results at the saturation field (2 kOe). This allows observing the dependence of the phase as a function of the Co magnetization 
direction relative to the drift velocity of the relaxing photo-carriers. An abrupt variation of the 
phase with the inversion of the magnetization direction has been 
explained with a simple model considering the inversion of the direction of Lorentz force acting on drifting 
photo-carriers as a consequence of the inversion of the Co magnetization (see the sketch in Fig. \ref{fig:fig3}(b))
explaining the effect in Figure \ref{fig:fig2}(b)). Independent experimental studies involving the analysis 
of the influence of magnetization direction on the characteristic relaxation time (back to equilibrium)  
shown in Figure \ref{fig:fig3}(a) support the above simplified model. 

Our observations suggest a further modification of the drift-diffusion equation when the metallic overlayer is 
made by a thin and narrow feromagnetic films \cite{Cascales2014}, where the resistance (R), effective 
capacitance (C) and inductance (L) of the metallic line becoming potentially magnetic field dependent, i.e. 
$L(H)\frac{d^{2}u(x_{0},t)}{dt^{2}}+R(H)\frac{du(x_{0},t)}{dt}+\frac{1}{C(H)}u(x_{0},t)=F(x_{0},t)$ where
$u(x,t)$ is the potential distribution, $x_0$ the laser spot position and $F(x_0,t)$ the 
electron-hole generation function. On one hand, the present experiments and simulations clearly show, in agreement with recent studies of stationary 
LPE by Wang et al. \cite{S_WANG2014}, that the magnitude of the LPE could be magnetically controlled 
through a magnetic field dependent resistance $R(H)$ originated from the lateral photoeffect and the 
anisotropic magnetoresistance (Figs. \ref{fig:fig4}(b)). On the other hand, our observation of a slow relaxation time
and dephasing dependent on the magnetic state indicate that in the present experimental conditions with an in-plane
magnetic field perpendicular to the 
ferromagnetic line structure, the main influence of the Cobalt magnetic state on the T-LPE dynamics
should be atributed to changes in the effective capacitance $C(H)$. Future experiments 
should investigate the influence of different (longitudinal) magnetic field configurations and explore the
practical aspects of tuning the PSD sensitivity by magnetic fields in Ferromagnetic/Oxide/Semiconductor structures.
\vspace{0.5cm}

We would like to acknowledge Jose Rodrigo and Laura Mart\'in for their help at the initial stage 
of the experiment as well as Ch. van Haesendonk
for the growth of the samples. We also acknowledge the support by the Spanish MINECO (MAT2012-32743),
and the Comunidad de Madrid through NANOFRONTMAG-CM (S2013/MIT-2850) and CCC-UAM (SVORTEX).

\end{document}